\documentclass[%
 reprint,
 amsmath,amssymb,
 aps,
]{revtex4-2}

\usepackage{xcolor}
\usepackage{hyperref}

\usepackage{tikz}
\usetikzlibrary{arrows.meta}

\definecolor{plum}{rgb}{0.36078, 0.20784, 0.4}
\definecolor{chameleon}{rgb}{0.30588, 0.60392, 0.023529}
\definecolor{cornflower}{rgb}{0.12549, 0.29020, 0.52941}
\definecolor{scarlet}{rgb}{0.8, 0, 0}
\definecolor{brick}{rgb}{0.64314, 0, 0}
\definecolor{sunrise}{rgb}{0.80784, 0.36078, 0}
\definecolor{lightblue}{rgb}{0.15,0.35,0.75}
\definecolor{carolina}{RGB}{153, 186, 221}
\tikzstyle{axisarrow} = [-{Latex[inset=0pt,length=5pt]}]

\begin{document}

\preprint{APS/123-QED}

\title{Prospects for observing chiral symmetry breaking in lepton colliders}

\author{Maurice H.P.M. van Putten$^{1,2}$ and Maryam A. Abchouyeh$^1$}
\email{mvp@sejong.ac.kr}
\affiliation{$^1$Physics and Astronomy, Sejong University, 209 Neungdong-ro, Seoul, South Korea,} 
\affiliation{$^2$INAF-OAS Bologna via P. Gobetti 101 I-40129 Bologna Italy,
Italy} 

\date{\today}

\begin{abstract}
Weak interactions in neutron $\beta$-decay exhibit parity violation through the preferential emission of right-handed antineutrinos. 
We identify this symmetry breaking with a reduction of phase space due to the small neutrino mass. During a brief interval of momentum exchange, 
a small mass neutrino puts the emission process close to the bifurcation horizon of Rindler space, doubly covered by Minkowski space ${\cal M}$ as dictated by the Equivalence Principle of general relativity. 
In the limit of arbitrarily small mass, this two-sheet covering effectively collapses into a single sheet, reducing the dimension of Dirac spinors from four to two, leaving neutrinos single-handed.  
This predicts a similar reduction to single-handed particle states in electrons created at TeV energies, which may be tested with the planned linear leptonic colliders. If confirmed, right-handed small mass neutrinos are expected to 
exist at sufficiently low energies.
\end{abstract}

\keywords{black holes -- symmetry}

\maketitle

\begin{figure}
\vskip-0.1in
\includegraphics[scale=0.4]{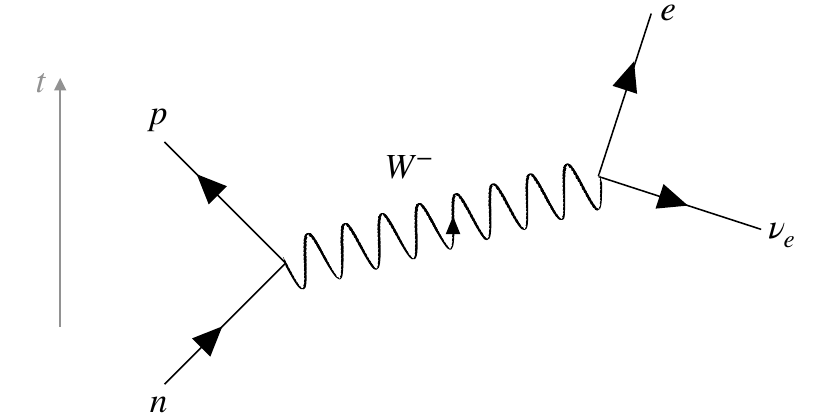} 
\caption{
$\beta$-decay of neutrons ($n$) exemplifies a brief instance of momentum exchange in the Standard Model, in the transformation to a proton ($p$) by changing a down to an up quark mediated by $W^-$. Its decay creates an electron ($e$) and an electron antineutrino ($\bar{\nu}_e$), 
preserving lepton number zero. Over the time-scale of creation of these leptons is described in a double cover in Minkowski spacetime by the Equivalence Principle of general relativity.}
\label{FIG_beta}
\end{figure}

{\em Introduction.} Neutrinos and their underlying physics play a crucial role in the Standard Model of particle physics, particularly in weak interactions, cosmology, and the identification of cosmic rays due to their unique properties
\citep{wor22}.
Notably, the Standard Model presents parity violation in weak interactions involving neutrinos of {\it small mass} \citep{lee56,wu57,zub20,tho13}.
A prototypical example is the $\beta$-decay of a neutron ($n$) into a proton $(p)$, electron and
antineutrino $\bar{\nu}_e$ (Fig. \ref{FIG_beta}),
\begin{eqnarray}
    n\rightarrow p + e^- +\bar{\nu}_e,
    \label{EQN_n}
\end{eqnarray}
where $\bar{\nu}_e$ is right-handed with lepton number $L=-1$. These interactions are commonly described by Feynman diagrams for given on-shell in- and output momentum states in Minkowski space ${\cal M}$. The mechanism of this chiral symmetry breaking giving rise to single-handedness for neutrinos is central to the Standard Model of particle physics and its extensions such as Supersymmetry (SUSY), whose origin remains rather mysterious.

In this {\it Letter}, we consider chiral symmetry breaking in (\ref{EQN_n}) as a consequence of small neutrino mass. 
Our starting point will be the conservation of momentum in the process of neutrino creation in Fig. \ref{FIG_beta} that, according to the Equivalence principle of general relativity, implies a double covering of
its world-line in ${\cal R} $ by hyperbolic world-lines in Minkowski spacetime ${\cal M}$. 
This follows some recent developments on the topology of gravitational collapse to black holes \citep{mvp2024}.
The double covering ${\cal M}$ of Rindler spacetime ${\cal R}$ in the comoving frame of the $\bar{\nu}_e$ is distinct from the conventional 
interpretation in a single covering of ${\cal R}$ by one wedge of ${\cal M}$, 
familiar from the definition of the Unruh-Davies vacuum \citep{ful73,dav75,unr76,tho84,tho85}. A distinction of double versus single covering of the observer's spacetime has also been considered in the Kruskal extension of Schwarzschild black holes \citep{tho84,tho22}.

\begin{figure*}[ht!]
\centering    \includegraphics[scale=0.65]{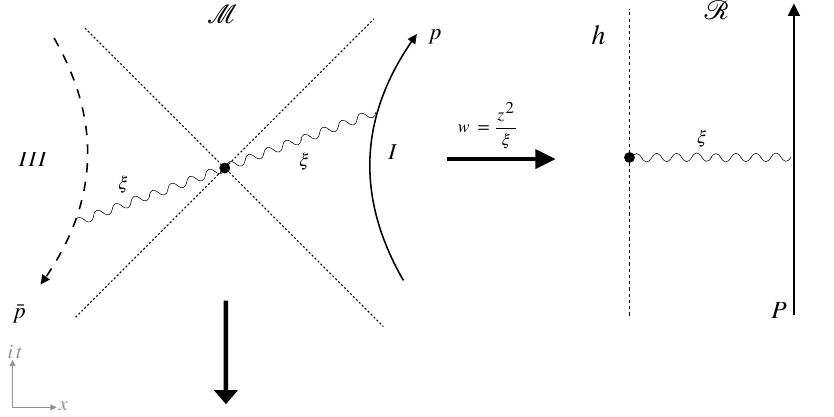}\\
\includegraphics[scale=0.65]{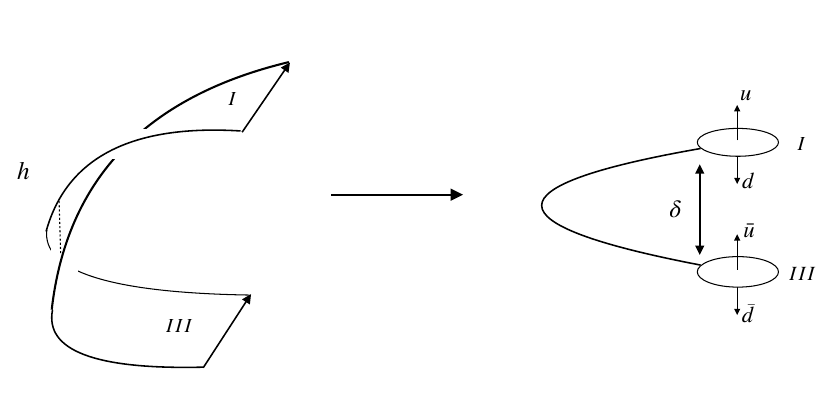}
\caption{(Top diagrams.) 
Double covering ${\cal M}$ of ${\cal R}$ by the map (\ref{EQN_w}), showing a Rindler (accelerated) 
particle $P$ in ${\cal R}$ and its pre-images $p$ (particle) and $\bar{p}$ (antiparticle) along hyperbolic trajectories in wedges I in III of ${\cal M}$ that propagate forward and, respectively, backward in time with respective lepton numbers $L=\pm 1$. The curvy line is the Lorentz invariant separation $\xi$ to the Rindler horizon $h$. 
(Lower diagrams.) The bifurcation horizon $h$ in ${\cal R}$ represents fold and a twist from wedge I to III on ${\cal M}$, equivalent to a M\"obius fold resulting in a two-sheet cover of ${\cal R}$ with wedge III facing up to the back of wedge I at some separation $\delta$.}
\label{FIG_spin1-1}
\label{FIG_spin1-2}
\label{FIG_spin1}
\end{figure*}

While a double covering supports the four degrees of freedom of fermions described by Dirac spinors, 
the limit of arbitrary small mass potentially introduces a degeneracy in collapse to a single-sheet. Formally, the zero-mass limit reduces the number of degrees of freedom by one-half from four to two, corresponding to the orientability of a single sheet. 

In \S2, the double covering representation of fermions is described. Upon close inspection, in \S3,
this change in topology is shown to leave only left-handed neutrinos and their right-handed antineutrinos to exist.
We summarize our interpretation with an outlook for future tests with the planned  {\it International Linear Collider} (ILC) and the {\it Compact Linear Collider} (CLIC) \citep{ILC19,sic20,bha20} in \S4.

{\it Double covering of lepton spinors.} The creation of $\bar{\nu}_e$ in (\ref{EQN_n}) occurs during a brief interval of momentum exchange about the origin -- the vertex of the light cone of the covering space ${\cal M}$ 
(Fig. \ref{FIG_beta} and Fig. \ref{FIG_spin1-1}, left) and equivalently about a bifurcation horizon $h$ in comoving Rindler space ${\cal R}$. This covering is described by the complex map 
\citep{mvp2024}
\begin{eqnarray}
    w=\frac{z^2}{\xi}.
    \label{EQN_w}
\end{eqnarray}
Here, $z$ is the world-line in Minkowski spacetime ${\cal M}$ and, $\xi=c^2/a$ parameterizes the Lorentz invariant distance to the origin in ${\cal M}$ at acceleration $a$ and, respectively, the Rindler horizon $h$ in ${\cal R}$ \citep{rin60} (see Fig. \ref{FIG_spin1}). 
Since $\xi$ is a distance, ${\cal R}(x,t)$ is a 1+1 spacetime and likewise for its covering space ${\cal M}(t,x)$.

At finite acceleration, the world-lines of particle $p$ and antiparticle $\bar{p}$ are hyperbolic extending in space-like separated wedges across the light cone of ${\cal M}$. Starting with $z=x+it\in {\cal M}(t,x)$, 
(\ref{EQN_w}) maps an antipodal pair 
of hyperbolic trajectories of uniformly accelerated particles $p$ and antiparticles ${\bar{p}}$ in ${\cal M}$ to a straight line in ${\cal R}$,
\begin{eqnarray}
z=\pm\xi \left(\cosh\lambda+i\sinh\lambda\right)\rightarrow 
w=\xi\left[1+i\sinh\left(2\lambda\right)\right]
\label{EQN_zw}
\end{eqnarray}
with $\lambda = a\tau/c$, where $\tau$ and $c$ are respectively, eigentime and velocity of light. (\ref{EQN_w}) and (\ref{EQN_zw}), hereby, show the double cover of the wold-line of an accelerating particle $P$ in ${\cal R}$ by a pair of hyperbolic world-lines in ${\cal M}$.

Since ${\cal M}$ is an orientable manifold of dimension two, wedges $I$ and $III$ each support two states of particles. 
For fermions, the spin states $(u,d)$ of $p$ up and down in $I$ ($(\bar{u},\bar{d})$ for $\bar{p}$ in $III$) can be identified with the two sides of ${\cal M}(t,x)$, and represented by a pair of spinors.
In (\ref{EQN_zw}), the extra factor of 2 in dependence on eigentime in $w\in {\cal R}$ implies evolution over a single period of a wave function of $P$ to signal a sign change in spinors along $z\in{\cal M}$, characteristic for observing spin-$\frac{1}{2}$ particles.
$P$ in ${\cal R}$ is hereby observed as a Dirac spinor by a pair of spinors of $p$ and $\bar{p}$ in ${\cal M}$.
By the same covering (\ref{EQN_w}), the light cone of ${\cal M}$ appears with antipodal symmetry in ${\cal R}$. In gravitational collapse starting from ${\cal M}$ pre-collapse, continuation of null-generators of its light cone shows the formation of event horizons $H$ in the topology $\mathbb{RP}^2$ with the Schwarzschild radius of the enclosed mass \citep{mvp2024}. 

\begin{figure*}
\center{
\vskip-0.0in
\includegraphics[scale=0.38]{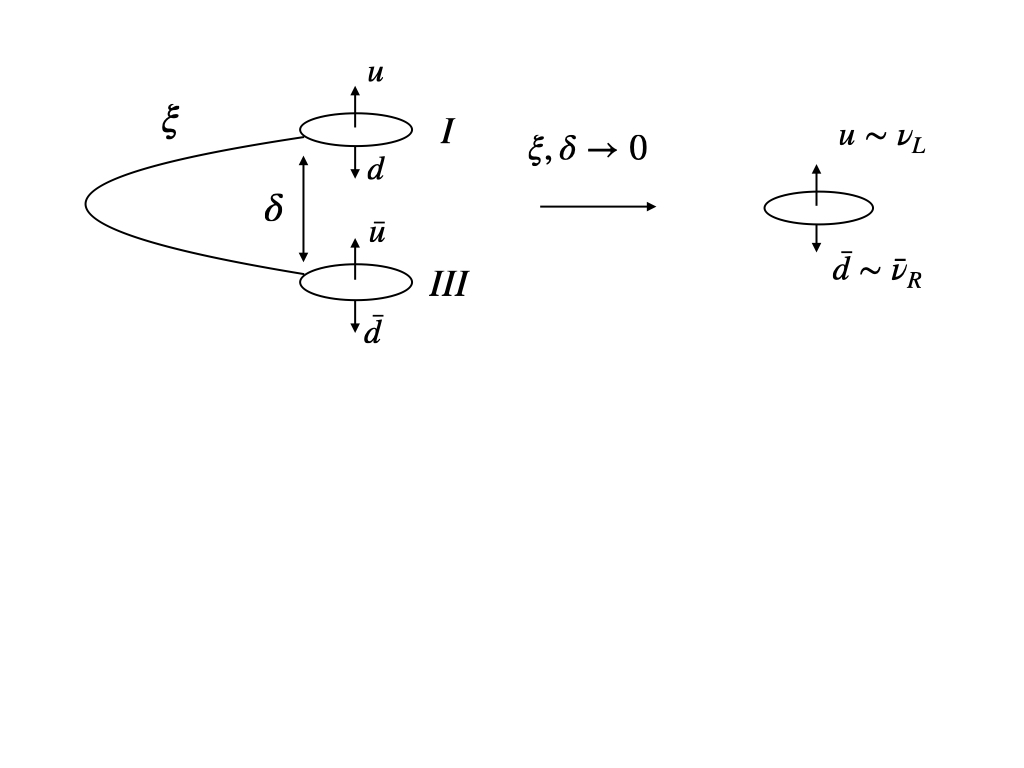} 
\vskip-2.2in
}
\caption{
In the limit of massless neutrinos $\xi\rightarrow0$, the trajectories of $p$ and $\bar{p}$ in Fig. \ref{FIG_spin1-1} approach the origin. (Right.) This reduces the double covering of ${\cal R}$ to a single sheet by coalescence of the two-sheets from wedge I and wedge III under (\ref{EQN_w}). As one sheet accommodates only two states of neutrinos,  we naturally identify 
$u=\nu_L$ and $\bar{\nu}_R=\bar{d}$, representing the opposite helicity states in the ultra-relativistic of small mass neutrinos.
}
\label{FIG_nu}
\end{figure*}

${\cal R}$ over the distance $\xi>0$ is covered by wedge I $(x>0)$ and wedge III $(x<0)$ of ${\cal M}(t,x)$. This effectively follows a M\"obius fold: a fold about the time axis with a M\"obius twist about the $x$-axis and conjugation of a spinor from III. 
If $x$-axis in ${\cal M}$ is right-handed, wedge I appears right-handed while wedge III appears left-handed in ${\cal R}$ - handedness changing under the fold but not the twist. 

Therefore, a straight but non-geodesic world-line in ${\cal R}$ is covered by hyperbolic orbits (\ref{EQN_zw}) in ${\cal M}$ of positive and negative energy particles $p,\ \bar p$ traversing forward and backward in time in wedges I and, respectively, III. The spinors of $p$ and $\bar{p}$ in ${\cal M}$ appear merged into the Dirac spinor of a Rindler particle $P$ in ${\cal R}$.
Applied to neutrinos, wedge I (III) naturally carries lepton number $L=1$ $(L=-1)$. 
In Fig. \ref{FIG_spin1-2}, orientability of ${\cal M}$ allows {\em face up} ($u$) and {\em face down} ($d$), while preserving the distinct lepton numbers in Dirac spinors in ${\cal R}$. 

For a given momentum exchange such as (\ref{EQN_n}), $\xi$ in Fig. \ref{FIG_spin1} scales with the mass of the particle by conservation of total momentum leaving a separation $\delta>0$ between the two sheets in ${\cal R}$. In what follows, we study the consequences in the limit of small mass.

{\it Small mass limit.} As illustrated in Fig. \ref{FIG_spin1-2}, a given momentum-exchange interaction with small mass is accompanied by a small $\xi$ by (\ref{EQN_w}).
Therefore, (\ref{EQN_n}) with small neutrino mass takes place close to the vertex of the light cone in ${\cal M}$, equivalently, $h$ in ${\cal R}$ (Fig. \ref{FIG_spin1-1}). It may be noted that the lifetime of $h$ equals the duration of the momentum interaction.

Conventionally, the distance $\xi$ to a Rindler horizin $h$ is astronomically large and effectively unobservable. For instance, at a nominal acceleration $g=9.8\,{\rm m\,s}^{-2}$, 
$\xi = {c^2}/{g} \simeq 
9\times10^{17}{\rm cm}\simeq 
0.36\,{\rm pc}$,
or about 1 lyr. In fact, in the outskirts of galactic disks, $h$ readily drops behind the Hubble horizon, where equality holds at the de Sitter scale of acceleration $a_{dS}=cH_0$ for a Hubble constant $H_0$ \citep{mvp2025}.

In contrast, consider the semi-classical approximation in the interaction of elementary particle. 
In $\bar p p$-collision at a non-relativistic energy 
$E_k\simeq (1/2)E_0\beta^2$, $0<\beta={v}/{c}\ll1$,
an inelastic collision extends over the size $d\simeq \lambda$ roughly given by the Compton wavelength $\lambda$, $2\pi/\lambda = mc/\hbar$.
During deceleration, we have 
$\xi \simeq {2\lambda}/{\beta^2}$, showing a proximity of $h$ in
\begin{eqnarray}
\frac{\xi}{\lambda} \simeq 
\frac{2}{\beta^2} \ll1.
\label{Example_32}
\end{eqnarray}
The resulting close proximity of the hyperbolic orbits of $p$ and $\bar{p}$ in ${\cal M}$ define a small separation $\delta$ shown in Fig. \ref{FIG_nu}. While any radiative corrections thereto are not known for massive particles, we next turn to the special case of neutrinos given their small mass.

The distance $\delta\simeq 2\xi$ between the hyperbolic trajectories $p$ and $\bar{p}\in {\cal M}$ sets a separation in the double covering ${\cal M}$ of ${\cal R}$, where $P$ in ${\cal R}$ is represented by a Dirac spinor $\left(\kappa^A,\bar{o}^A\right)$ with $\kappa^A$ from wedge I and $o^A$ from wedge III in ${\cal M}$.
In a given momentum-exchange interaction, the massless limit hereby takes us to $\delta\rightarrow0^+$ as $\xi\rightarrow0^+$.

Fig. \ref{FIG_nu} shows the limit $\delta\rightarrow0^+$ to reduce the covering of ${\cal R}$ to a single sheet, by coalescence of face-up of wedge III and face-down of wedge I.
This reduced covering leaves $P$ to observe effectively a Weyl spinor $\iota^A$  possessing the states $u$/I ($L=1)$ and $\bar{d}$/III $(L=-1)$. 

Observations on (\ref{EQN_n}) show the production of only right-handed but not left-handed antineutrinos over a broad energy range of order 1\,MeV. We identify these observations with $\delta$ sufficiently small to effectively reduce to $\delta=0$ in Fig. (\ref{FIG_nu}).
With $\delta \rightarrow 0^+$ equivalent to high energy scales compared to rest-mass energy at large Lorentz factors, and considering the orientability of each single sheet, neutrinos are, as expected, observed to be single-handed. This suggests a window $0<\delta=\delta_{\nu}<\delta_c$ 
for some transition scale $\delta_c>0$, where the effectively one-sheet cover applies in current observations. 
Based on (\ref{EQN_w}), while unknown, $\delta_c$ should be sufficiently small such that the more massive electrons and positrons preserve their four distinct states consistent with existing experimental energies, corresponding to $\delta=\delta_e \gg \delta_c$. 

{\it Conclusions and outlook.} We revisit the problem of single-handed neutrinos in (\ref{EQN_n}) by the topology of covering spaces in (\ref{EQN_w}) {\it during momentum exchange interactions}. Based on the Equivalence Principle of general relativity, the world-line of the antineutrino in its comoving Rindler frame of reference ${\cal R}$ is covered by a pair of hyperbolic orbits (\ref{EQN_zw}) in ${\cal M}$, following the map (\ref{EQN_w}).

Ordinarily, this presents a double covering by (\ref{EQN_w}). However, in the limit of small mass or sufficiently high-energy scales compared to rest mass energy, this covering potentially degenerates to a single cover by the associated large accelerations for a given momentum exchange. 
Such limit reduces the phase space of states from four to two.
In the case of (\ref{EQN_n}), this leaves neutrinos single-handed.
Conversely, this suggests that right-handed 
neutrinos conceivably appear at sufficiently low energies, below the conventional energy scales of $E_{\bar{\nu}_e}\sim 1\,$MeV. However, the detection of neutrinos below this energy scale is notoriously difficult. 

Instead, our topological origin of single-handedness may be tested with the planned high-energy lepton collider experiments. With a larger mass $m_e$, electron-positron interactions may share 
similarly small values $0<\delta_e<\delta_c$ at energies
\begin{eqnarray}
    E_e\gtrsim \left(\frac{m_e}{m_\nu}\right)\,E_{\bar{\nu}_e},
    \label{EQN_Ee}
\end{eqnarray}
where $E_{{\nu}_e}$ is the energy scale at which antineutrinos in (\ref{EQN_n}) are observed. 
Given $E_{\bar{\nu}_e}\sim 0.1-1\,$MeV in (\ref{EQN_n}) 
and $m_\nu c^2\sim 0.1-1$\,eV, the scaling (\ref{EQN_Ee}) points to electron energies on the order of 0.1-10\,TeV, which is expected to be accessible with the planned ILC and CLIC experiments. 
At these energies, we may see a natural selection to single-handedness in the creation of electron-positron based on universality of leptonic interactions at high energies, where electrons and neutrinos are expected to share the same symmetries. 

If confirmed in ILC-CLIC experiments, a change in chiral symmetry of electrons across sufficiently high energies (\ref{EQN_Ee}) would provide compelling support for a reverse change for  neutrinos at sufficiently low energies, otherwise inaccessible by direct observation. 

{\bf Acknowledgments.} 
This research is supported, in part, by 
NRF grant No. RS-2024-00334550.


\end{document}